# Computational Realization of Popping Impinging Sprays of Hypergolic Bipropellants by a Eulerian-Lagrangian Approach

Jinyang Wang[1,2], Kai Sun[1]*, Tianyou Wang[1], and Peng Zhang[2]*

[1] State Key Laboratory of Engines, Tianjin University, Tianjin 300350, P.R. China

[2] Department of Mechanical Engineering, City University of Hong Kong, Kowloon Tong, Kowloon, 999077, Hong Kong

\* Authors to whom correspondence should be addressed:

sunkai@tju.edu.cn and penzhang@cityu.edu.hk

## ABSTRACT

This work adopts a Eulerian-Lagrangian approach to numerically simulate the spray impingement of MMH (Monomethyl hydrazine)/NTO (nitrogen tetroxide), which are prevalent rocket engine bipropellants for deep space missions and satellite orbital maneuvers. The emphasis of the work is to computationally realize the popping phenomenon and to study its parametric dependence on liquid and gas-phase reaction rates. The liquid-phase reaction of MMH/NTO is realized based on the extended spray equation, incorporating the additional independent variable, propellant mass fraction, to account for the mixing of droplets. The spray popping can be computationally reproduced over wide ranges of Damköhler numbers for both liquid- and gas-phase reactions. Furthermore, the computational results have been validated through qualitative comparison with experimental images and quantitative comparison with experimental frequencies. The present results verify our hypothesis that the heat release from the liquid-phase reaction enhances the evaporation of MMH and NTO so that the intense gas-phase reaction zone around the spray impingement point periodically separates the MMH and NTO impinging sprays to cause the popping phenomenon. Furthermore, it was found that the popping phenomenon can be suppressed by reducing the Damköhler numbers of liquid-phase reaction and therefore to suppress the evaporation of the propellants. This work is believed to provide valuable understanding for avoiding the off-design popping phenomenon that may reduce combustion efficiency and increase the risk of combustion instability in rocket engines.

# 1. Introduction

Hypergolic bipropellants are widely used in liquid rocket engines for satellite orbital maneuvers, deep space missions, attitude control, and lifetime termination due to their high combustion efficiency, rapid response, and self-ignition [1–3]. In contrast to non-hypergolic bipropellants, hypergolic bipropellants do not require a separate ignition system even in harsh environments with low temperatures and pressures. This makes the combustion process easy and allows multiple cold starts, rendering simple and reliable engine operation.

Liquid rocket engines generally employ impingement-type injectors[2] to utilize hypergolic bipropellants. Two propellants are injected separately into the combustion chamber and impinge with each other at high velocities, resulting in the primary atomization of the liquid propellant streams. At the same time, the two propellants encounter and mix in the forms of liquid sheets and droplets. Due to the hypergolic nature of the propellants that ignite automatically on contact with each other, liquid phase reactions occur in the mixed liquid sheets and droplets, followed by the evaporation and gas-phase reactions of the bipropellants[2,4].

As early as the 1960s, NASA conducted a series of experimental studies[5–11] of the monomethylhydrazine (MMH, $CH_3NHNH_2$) - dinitrogen tetroxide (NTO, $N_2O_4$) self-igniting thruster combustion system to understand the flow and combustion of hypergolic bipropellants. Among various identified combustion modes[8], the popping mode has attracted significant attention, as schematized in Fig. 1 According to the engine design, MMH and NTO would encounter, mix, and react, and the combustion is in the mixed mode, as shown in Fig. 1(a)-(b). However, MMH and NTO may separate from each other, as shown in Fig. 1(c), which is speculated to be related to the intense reactions. If they remain separated for a prolonged duration, the combustion is in the mode of reaction stream separation (RSS) with deteriorated combustion efficiency. If MMH and NTO re-impinge again, as shown in Fig. 1(d), and repeat the separation-impingement periodically, the combustion is in the mode of "popping", which may increase the risk of combustion instability and reduce combustion efficiency.

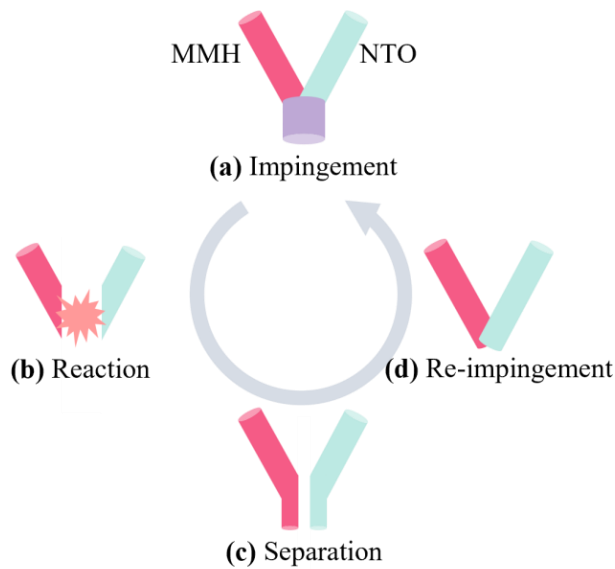

**FIG. 1.** Schematic diagram of popping phenomenon of hypergolic propellant impingement of MMH/NTO. (a) The impingement and mixing of MMH and NTO. (b) Intense reactions at the impingement point. (c) The two propellant streams are separated to impede the contact between MMH and NTO. (d) Re-impingement of MMH and NTO streams.

In recent years, Dennis et al.[12] investigated critical ignition thresholds by varying the jet diameter and total propellant mass flow rate at constant oxidizer-fuel and momentum ratios. They observed good ignition, transition region, and no ignition, and their attention was on the conditions of reaction stream separation. Similarly, Tani et al.[13] studied the jet and flame behavior near the impact point of hypergolic bipropellants and observed that the MMH and NTO jets were separated by the flame in the RSS mode. In the highly reactive hypergolic propellant systems, Houseman et al.[7] developed two theoretical models for low-pressure and high-pressure conditions respectively. The low-pressure model is based on the assumption that the jet collision point reaches the bubble-point temperature and is dominated by liquid-phase reactions. In the high-pressure model, the gas reactions are assumed to be sufficiently fast to maintain an air film between the impinging streams, causing the jets to separate and no longer contact each other. Lawver[8] summarized that the RSS is more likely to occur at high propellant temperatures, large injector sizes, and high chamber pressures. For the popping mode, previous studies[14–16] observed periodic explosions near the point of impingement point experimentally and found that the mechanism is similar to that of RSS. Houseman et al.[14,17] assumed that the popping results from the gas film's cyclic breakdown and pointed out the importance of the liquid-phase contact time, liquid-phase mixing, and liquid-phase reaction.

Zhang et al.[18–20] studied the hypergolic droplet collision, which is more consistent with the actual process in engines by eliminating the influence of the "wall effect" in the conventional drop tests. The impact parameters, droplet sizes, and eccentricity distances were considered for their influence on droplet-droplet mixing, reaction, and ignition. According to their studies, the ignition and combustion process of the colliding bipropellant droplets can be divided into three stages [18,19,21]: the first is the liquid-phase reaction, which generates heat to enhance vaporization; the second is the mixing and combustion of propellant vapors; and the third is the intense gas-phase reaction. The fast liquid phase reaction in the first phase is the key to the subsequent ignition and combustion processes in the gas phase. These findings are useful to our understanding of the fundamental microscopic processes of impingement combustion and the liquid reaction of hypergolic propellants[22].

However, the existing experimental analyses and theoretical assumptions of RSS and popping modes lack direct evidence, such as temperature fields and flow fields. To further understand the impingement and combustion processes of hypergolic propellants, numerical simulation might be an effective approach. Due to the complex physical-chemical processes[23], numerical simulation studies of MMH-NTO propellants impingement and combustion are much scarcer compared to extensive experimental research.

Furthermore, most computational studies only consider the gas-phase reactions of MMH/NTO in the combustion process. Wei et al.[24] extended the traditional non-coupled spray model to account for the coupling between the injection process and the surrounding gas field. This coupled model can more accurately predict key spray characteristics, such as the mass flux and mixture ratio distributions in quiescent air, and serves as the first step for complex simulations of hypergolic propellant coupling impingement and combustion. Lee et al. [25] conducted a computational study on a small MMH/NTO bipropellant liquid rocket engine, examining the effects of wall-cooling fuel fraction on propellant combustion flow, combustion efficiency, and wall temperature. Ohminami[26] simulated the combustion flow inside a film-cooled bipropellant thruster using a reduced mechanism of gas-phase reactions, focusing on mass fraction, temperature distribution, and wall heat flux. Lee[27] utilized a global kinetic reaction model of the actual combustion process of MMH-NTO to investigate more realistic plume behavior. Chu et al.[28] studied tangential combustion instability in an MMH/NTO rocket combustor, simplifying the problem by neglecting liquid-phase reactions. Wei et al.[29] attempted to implement liquid-phase reactions of MMH-NTO by establishing a one-step finite-rate liquid-phase hypergolic reaction model. Their simulation showed that the hypergolic liquid-phase reaction plays a crucial role in the igniting process of the thruster.

Despite the above computational efforts, understanding the influence of liquid-phase reactions on MMH-NTO ignition is still far from being complete[29,30], particularly on the hypergolic combustion modes of RSS and popping. To computationally realize liquid-phase reactions, it is necessary to develop a method for describing the mixing process of different propellants. Our previous studies[31,32] used the computational framework of Eulerian-Lagrangian spray simulations, introducing an additional independent variable, the propellant mass fraction $\varphi$, into the droplet probability distribution function.

This approach enables the simulation of dual-fuel spray impingement and liquid-phase mixing. We systematically studied the effects of various fuel properties (fuel density, latent heat of evaporation, and vapor pressure) and injection parameters (injection pressure and angle) on spray behaviors, liquid-phase mixing, mixture concentration distribution, and gas-phase ignition characteristics. These results form the foundation of the present study on modeling the liquid-phase reactions of MMH/NTO to computationally realize the popping phenomenon for the first time in literature, providing a more direct analysis of flow field information to elucidate physics and verify conjectures. This deepens our understanding of the combustion mechanisms of hypergolic bipropellants, contributing to the design of liquid rocket engines to improve combustion efficiency and stability.

## 2. Computational Methods

### 2.1. Eulerian-Lagrangian approach based on spray equation

This computational work adopts the Eulerian-Lagrangian approach to numerically simulate the MMH/NTO spray impingement and combustion process based on the KIVA-3v open-source simulation platform[33–35]. Our previous studies[31,32] and many other researches[36–38] have fully demonstrated the reliability of this approach.

The computational fame uses the Eulerian method to describe the flow of gas. The continuity equation for the *m*-th species is given by

$$\frac{\partial \rho_m}{\partial t} + \nabla \cdot (\rho_m \mathbf{u}) = \nabla \cdot \left[\rho D \nabla \left(\frac{\rho_m}{\rho}\right)\right] + \dot{\rho}_m^c + \dot{\rho}^s \delta_m, \quad (1)$$

where $\rho_m$ is the mass density of species *m*, $\rho$ is the total mass density, **u** is the velocity vector of fluid, *D* is a single diffusion coefficient of Fick's Law diffusion. The source terms $\dot{\rho}_m^c$ and $\dot{\rho}^s$ are chemistry and spray, respectively. $\delta_m$ is the Dirac delta function of the species *m* in spray droplets. Sum Eq. (1)

overall species *m* to obtain the total fluid density equation

$$\frac{\partial \rho}{\partial t} + \nabla \cdot (\rho \mathbf{u}) = \dot{\rho}^s . \tag{2}$$

The momentum equation for the mixture of fluid is given by

$$\frac{\partial \rho \mathbf{u}}{\partial t} + \nabla \cdot (\rho \mathbf{u}\mathbf{u}) = -\frac{1}{a^2}\nabla p - A_0 \nabla\left(\frac{2}{3\rho k}\right) + \nabla \cdot \boldsymbol{\sigma} + \mathbf{F}^s + \rho \mathbf{g}, \tag{3}$$

where $p$ is the pressure of the fluid, and $a$ is the dimensionless quantity of the Pressure Gradient Scaling (PGS) method. $A_0$ is zero or unity, respectively, when in laminar flow calculations or when using one of the turbulence modes. $\boldsymbol{\sigma}$ is the viscous stress tensor, $\mathbf{F}^s$ is the rate of momentum gain per unit volume by spray, $\mathbf{g}$ is the specific body force and constant assumed.

The energy equation is given by

$$\frac{\partial (\rho I)}{\partial t} + \nabla \cdot (\rho \mathbf{u}I) = -p\nabla \cdot \mathbf{u} + (1 - A_0)\boldsymbol{\sigma}:\nabla \mathbf{u} - \nabla \cdot \mathbf{J} + A_0 \rho \varepsilon + \dot{Q}^c + \dot{Q}^s, \tag{4}$$

where $I$ is the specific internal energy (excluding chemical energy). $\mathbf{J}$ is the heat flux vector, which is the sum of the contributions from heat conduction and enthalpy diffusion. The source terms $\dot{Q}^c$ and $\dot{Q}^s$ represent chemical heat release and spray interactions, respectively.

Due to the dynamics of droplets in the spray and their interaction with the gas, which are highly complex coupled problems, the Lagrangian method is utilized to describe this process. Specifically, the particle method[39] is used to solve the spray equation[40], which statistically describes the evolution of droplet distribution function (DDF) in a high-dimensional phase space for describing the discrete droplets.

$$f(\mathbf{x}, \mathbf{v}, r, T_d, y, \dot{y}, t) d\mathbf{x}\, d\mathbf{v}\, dr\, dT_d\, dy\, d\dot{y} \tag{5}$$

which represents the probable number of droplets per unit volume at time $t$ with locations in the range $(\mathbf{x}, \mathbf{x} + d\mathbf{x})$, velocities in the range $(\mathbf{v}, \mathbf{v} + d\mathbf{v})$, radii in the range $(r, r + dr)$, temperatures in the range $(T_d, T_d + dT_d)$, deviations in the range $(y, y + dy)$, deviation rates in the range $(\dot{y}, \dot{y} + d\dot{y})$, To simulate the actual collision and combustion of droplets during the spray impingement, additional independent variables, the propellant mass fractions $\varphi_i$ ($i = 1,2 \ldots, M$), are added to the DDF to represent the mass fraction of $M$ chemical components in liquid phase, and are defined by

$$\varphi_i = m_i / \sum_i^M m_i \tag{6}$$

where $m_i$ is the mass of the $i$-th chemical components. Apparently, we have $0 \leq \varphi_i \leq 1$ and $\sum_i^M \varphi_i = 1$ as the constraints of $\varphi_i$. In this paper, we consider only two chemical components, namely the bipropellants MMH and NTO, rendering $M = 2$ and only one independent propellant mass fraction denoted by $\varphi$ without a subscript. When $\varphi = 0$ and $\varphi = 1$, the droplets are pure NTO and MMH, respectively. It should be noted that, although a simplified mode of $M = 2$ was used in the present work, the formulation can be easily extended to arbitrary $M$ because the solution of the spray equation is not sensitive to the dimension of the DDF, to be expatiated shortly. This provides a solid foundation for the subsequent realization of liquid-phase reactions.

The newly expanded DDF containing the information on propellant mass fraction can be expressed as

$$f(\boldsymbol{x}, \boldsymbol{v}, r, T_d, y, \dot{y}, \varphi, t) d\boldsymbol{x} \, d\boldsymbol{v} \, dr \, dT_d \, dy \, d\dot{y} \, d\varphi, \tag{7}$$

where the additional dimension is the propellant mass fraction in the range $(\varphi, \varphi + d\varphi)$. The time evolution of DDF is governed by the extended spray equation,

$$\frac{\partial f}{\partial t} + \boldsymbol{\nabla}_x \cdot (f\boldsymbol{v}) + \boldsymbol{\nabla}_v \cdot (f\boldsymbol{F}) + \frac{\partial}{\partial r}(fR) + \frac{\partial}{\partial T_d}(f\dot{T}_d) + \frac{\partial}{\partial y}(f\dot{y}) + \frac{\partial}{\partial \dot{y}}(f\ddot{y}) \\ + \frac{\partial}{\partial \dot{\varphi}}(f\dot{\varphi}) = \dot{f}_{coll} + \dot{f}_{bu}, \tag{8}$$

where the quantities $\boldsymbol{F}, R, \dot{T}_d,$ and $\ddot{y}$ are the time change rates of velocity, radius, temperature, and deviation rate, respectively. The source terms of droplet collision and breakup can be represented by

$$\dot{f}_{coll} = \frac{1}{2} \iint f(x, v_1, r_1, T_{d_1}, y_1, \dot{y}_1, \varphi_1, t) \, f(x, v_2, r_2, T_{d_2}, y_2, \dot{y}_2, \varphi_2, t) \pi (r_1 + r_2)^2 \, |v_1 - v_2|$$
$$\{\sigma(v, r, T_d, y, \dot{y}, \varphi, v_1, r_1, T_{d_1}, y_1, \dot{y}_1, \varphi_1, v_2, r_2, T_{d_2}, y_2, \dot{y}_2, \varphi_2)$$
$$-\delta(v - v_1)\, \delta(r - r_1)\delta(T_d - T_{d_1})\delta(y - y_1)\delta(\dot{y} - \dot{y}_1)\delta(\varphi - \varphi_1)$$
$$-\delta(v - v_2)\, \delta(r - r_2)\delta(T_d - T_{d_2})\delta(y - y_2)\delta(\dot{y} - \dot{y}_2)\delta(\varphi - \varphi_2)\}$$
$$dv_1 \, dr_1 \, dT_{d_1} \, dy_1 \, d\dot{y}_1 \, d\varphi_1 \, dv_2 \, dr_2 \, dT_{d_2} \, dy_2 \, d\dot{y}_2 \, d\varphi_2 \quad (9)$$

$$\dot{f}_{bu} = \int f(x, v_1, r_1, T_{d_1}, y_1, \dot{y}_1, \varphi_1, t)\, \dot{y}_1 B(v, r, T_d, y, \dot{y}, \varphi, v_1, r_1, T_{d_1}, \dot{y}_1, \varphi_1, x, t)$$
$$dv_1 \, dr_1 \, dT_{d_1} \, dy_1 \, d\dot{y}_1 \, d\varphi_1, \quad (10)$$

where the subscripts 1 and 2 are properties of droplets 1 and 2. The functions $\sigma$ and $B$ represent the transition probabilities for collision and breakup, respectively. In this work, we assumed that droplet collision and breakup do not depend on the propellant mass fraction $\varphi$. In reality, the mixing of droplet of different propellants may change the physical properties of the merged mass and in turn the subsequent collision and breakup.

The particle method used to solve Eq. (8) is based on the discrete representation of the DDF,

$$f_N = \sum_{p=1}^{N} N_p \, \delta(x - x_p)\delta(v - v_p)\, \delta(r - r_p)\delta(T_d - T_{d_p})\delta(y - y_p)\delta(\dot{y} - \dot{y}_p)\delta(\varphi - \varphi_p) \quad (11)$$

where each computational parcel $p$ represents $N_p$ droplets with identical locations and properties. That the computational parcels do not represent the real physical droplets[41] is a significant difference from the Lagrangian Particle Tracking (LPT) method[42]. For the real physical droplets, the evolution of particle location and properties due to droplet/gas interaction and droplet collision is determined by a set of ordinary differential equations. In fact, they are related only by the requirement that they must have the same conditional expectations according to the principle of statistical equivalence[43,44]. Because N is the number of computational particles and can be a few orders of magnitude smaller than the number of droplet parcels used in the LPT method, the computational cost of the DDF approach is significantly smaller than that of the LPT method.

**2.2. Sub-models of spray and combustion**

Fig. 2 illustrates the sub-models of droplet breakup and collision in the Euler-Lagrangian computational framework based on the extended spray equation. As shown in Fig. 2(a), the breakup model used in this work is derived from the TAB model[45]. We assumed that all the daughter droplets with different radii formed by the breakup have the same propellant mass fraction $\varphi$ as their mother droplet. The O'Rourke collision model[46] is used for droplet collisions, as shown in Fig. 2(b). When the collision outcome is coalescence, the mass and composition of the merging droplet, denoted by the subscript $k$, are calculated based on the mass and composition of the two colliding droplets, denoted by subscripts $i$ and $j$.

$$\varphi_k = \frac{m_i \varphi_i + m_j \varphi_j}{m_i + m_j} \tag{12}$$

Here we assumed rapid coalescence and inter-mixing of droplet for simplicity. In reality, droplets take a finite time to complete merge and mixing[22], and this must be considered in future work for more sophisticated models. When the collision outcome is the temporary coalescence followed by stretching and separation, the propellant mass fractions of the separated droplets are the same as before the collision. Again, this model assumes that there is negligible inter-mixing within the temporarily merged droplet. The complex dynamics of internal droplet mixing and its dependence on the Marangoni effect[47,48] and size ratio effects are not fully considered in this study but will be further investigated in future research.

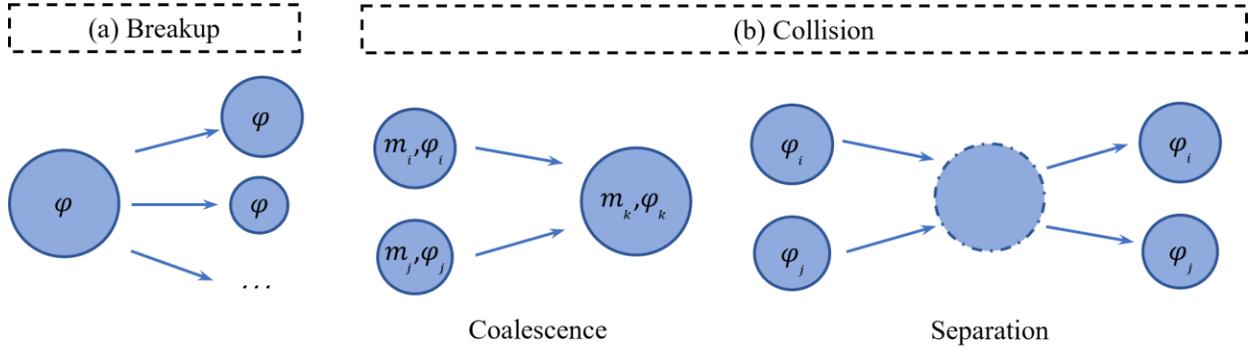

**FIG. 2.** Schematic of sub-models of droplet breakup and collision: (a) Breakup of single droplet. (b) Collision for coalescence and separation of binary droplets. $\varphi$ represents the mass fraction of MMH in droplets. The subscripts $i$, $j$, and $k$ denote different droplets.

The evaporation model adopted by this work is the Spalding model[49], a single-droplet evaporation model that ignores the multicomponent effects. In reality, the liquid-phase mass diffusion is crucial in the evaporation process of multicomponent mixed droplets and may result in micro-explosions[50,51]. Modeling multicomponent droplet evaporation is still a problem incompletely solved and will be considered in our future work. Consequently, we assumed that MMH and NTO vaporize independently in the mixed droplet, and their time-dependent masses can be calculated by

$$m_F(t+dt) = m_F(t) - v_F dt \tag{13}$$

$$m_O(t+dt) = m_O(t) - v_O dt \tag{14}$$

where $m_F(t)$, $m_O(t)$ and $m$ represent the mass of MMH, NTO, and total droplet at the current time, $m_F(t+dt)$ and $m_O(t+dt)$ denote the mass of MMH and NTO after an interval time $dt$, and $v_F$ and $v_O$ are their evaporation rates, respectively.

Since the detailed validated chemical reaction mechanism of MMH-NTO is still under development, and, more importantly, the present study focuses on the dynamical characterization of the spray in a combustion environment, we adopted simplified reaction mechanisms to describe the finite-rate combustion process of MMH-NTO in both gas-phase and liquid-phase. It will be seen shortly that such simplified mechanisms enable the detailed characterization of the hypergolic ignition by using Damköhler numbers for both gas-phase and liquid-phase reactions.

The gas-phase one-step reaction mechanism is $4MMH + 5NTO \rightarrow 9N_2 + 12H_2 + 4CO_2$ with the gas-phase reaction rate is $\omega_g = k_g[MMH][NTO]^{1.25}$. The reaction rate constant is given by $k_g = A_g e^{-E_{a_g}/RT_g}$, where $A_g$ is the pre-exponential factor of gas-phase reaction, $E_{a_g}$ is activation energy of gas- phase reaction, $T_g$ is the temperature of the gas phase in the reaction[52,53]. The liquid phase reaction mechanism is $MMH(l) + NTO(l) \rightarrow Product(g)$ with the reaction rate $\omega_l = k_l[MMH][NTO]$. The liquid-phase reaction rate constant $k_l = A_l e^{-E_{a_l}/RT_l}$, where $A_l$ is the pre-exponential factor of liquid-phase reaction, $E_{a_l}$ is activation energy of liquid-phase reaction, $T_l$ is the temperature of the liquid phase in the reaction. For the uncertainty analysis, we considered the reaction rate constants for the gas and liquid phases in the range of $k_g \sim 10^9 - 10^{14}$ $(cm^3\ mol^{-1})^{1.25}\ s^{-1}$ and $k_l \sim 10^{13} - 10^{19}$ $cm^3\ mol^{-1}\ s^{-1}$, respectively. The evolution of the mass of the MMH and NTO within a mixed droplet due to reactions is given by

$$m_F(t + dt) = m_F(t) - \omega_l dt \tag{15}$$

$$m_O(t + dt) = m_O(t) - \omega_l dt \tag{16}$$

Modelling the gas-phase reactions has been described in great detail in the literature[32] and will not be repeated here.

## 2.3 Computational Specifications

In this work, the computational domain is 100mm × 100mm × 100mm, closely resembling the dimensions of the experimental combustion chambers[13,54]. A detailed study of mesh independence was performed in our previous work[32,55], which examined the dependence of the spray mixing and combustion process on droplet collision dynamics. The numerical results are not sensitive to the mesh size when the mesh size is smaller than 2mm × 2mm × 2mm. Therefore, we selected the mesh size of 2 mm × 2 mm × 2 mm, which is consistent with our previous studies[31,32,55] and those by Kim et al.[56] and Suo et al.[57]. The computational setup is illustrated in Fig. 3. The injectors are positioned on the two sides around the center of the top surface on the cubic computational domain with injector distance $L$. The impingement point is the intersection of the velocity directions of the two sprays, determined by the impingement angle $\theta$. The region and plane used for analyzing the spray characteristics at the impingement point are an 8mm × 8mm × 10mm cubic box and an 8mm × 10mm rectangular Plane 1, positioned at the center of the box. There is a significant variation in the number of droplets within this cubic box at the moments of spray impingement and separation, clearly reflecting the characteristic features of the popping phenomenon.

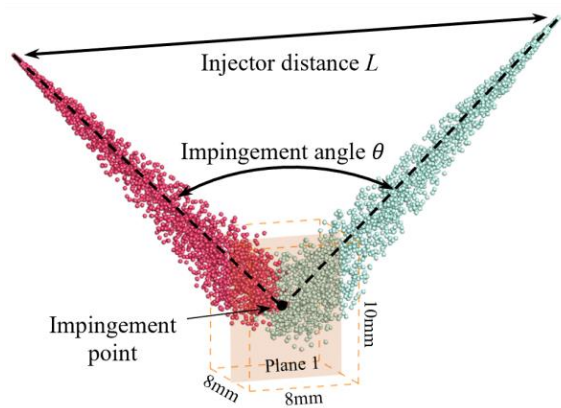

**FIG. 3.** Problem description of MMN/NTO spray impingement.

**2.4 Damköhler Number Characterization**

Both the spray flow and the chemical reactions in liquid- and gas-phase play an important role in the hypergolic ignition process. Therefore, the Damköhler number is the appropriate dimensionless parameter used to characterize the ignition processes[52,53], and it is defined by

$$Da = \tau_f/\tau_c \qquad (17)$$

where the characteristic flow time is $\tau_f = D/V$, $D = 0.01\ cm$ is the orifice diameter and $V = 220\ m/s$ is the spray injection velocity; $\tau_c$ is the characteristic chemical reaction time. Owing to the one-step reaction mechanisms adopted by the study, the Damköhler numbers for gas-phase and liquid-phase reactions are defined as

$$Da_g = k_g C_g^{n_g-1} D/V \qquad (18)$$

$$Da_l = k_l C_l^{n_l-1} D/V \qquad (19)$$

where $n_g = 2.25$ and $n_l = 2$ are reaction orders of gas-phase and liquid-phase reactions, respectively. The constants $C_g$ and $C_l$ are given by $C_g = \frac{1}{V_m}$ and $C_l = \frac{1}{2}\left(\frac{1}{V_F} + \frac{1}{V_O}\right)$, where $V_m = 22.4\ L/mol$ represents the molar volume of gas under standard conditions, and $V_F$ and $V_O$ represent the molar volumes of MMH and NTO in liquid phase, respectively.

## 3. Results and Discussion

### 3.1 Phenomenology and Regime Nomogram of Popping and Penetration Modes

Fig. 4(a) illustrates a typical popping case in the MMH-NTO combustion process. The key physical parameters of the benchmark case are as follows: $Da_g$=1.42×10², $Da_l$=2.84×10¹, impingement angle $\theta = 90°$, and injector distance $L = 4\ mm$. The dimensionless time is defined as $\tau = t/\tau_f$, where $t$ is the physical time. MMH and NTO sprays impinge with each other at $\tau = 0.0$. Subsequently, they are separated at $\tau = 44.0$ but come into impinge again at $\tau = 131.9$. This cyclical process of impingement - separation - re-impingement was experimentally observed and referred to as the popping phenomenon [16], as shown in Fig. 4(b). Although the controlling parameters are not the same as those of the experiment, the simulation results effectively capture the characteristic popping phenomenon. This is the first time that the popping phenomenon in hypergolic combustion is computationally realized by using the present approach.

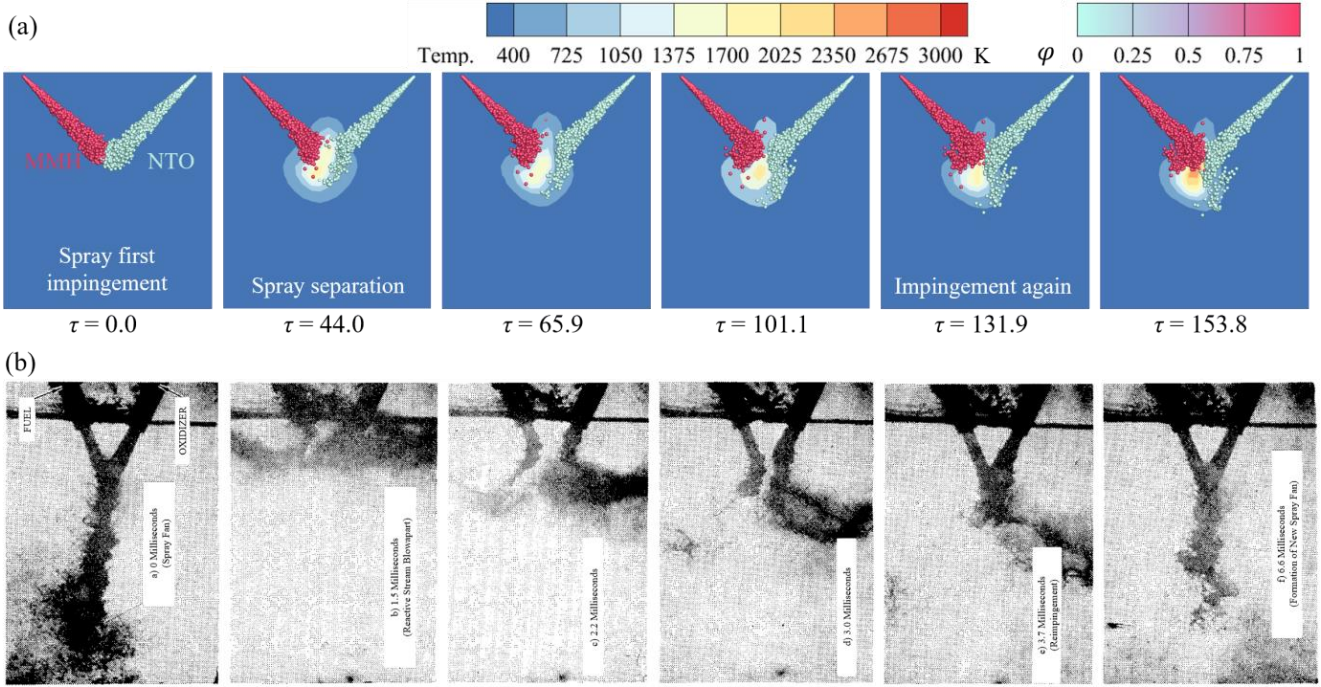

**FIG. 4.** (a) Computational benchmark case illustrates the popping phenomenon with $Da_g$=1.42×10$^2$, $Da_l$=2.84×10$^1$, $\theta = 90°$, and $L = 4\ mm$. The contour indicates gas field temperature and the spray is colored by the propellant mass fraction $\varphi$ for distinguishing MMH and NTO. (b) Experiment of popping phenomenon[16].

Another typical case is spray penetration. Compared to the benchmark case, the penetration case differs only in the smaller $Da_l$=2.84×10$^{-3}$, resulting in a slower liquid-phase reaction. After impingement, MMH and NTO penetrate each other rather than separate, as shown in Fig. 5(a). This phenomenon closely resembles the penetrating mode observed experimentally[58], as shown in Fig. 5(b). Higher injection pressures were used in experiments to achieve greater spray velocities and smaller Damköhler numbers, resulting in sprays penetrating through each other.

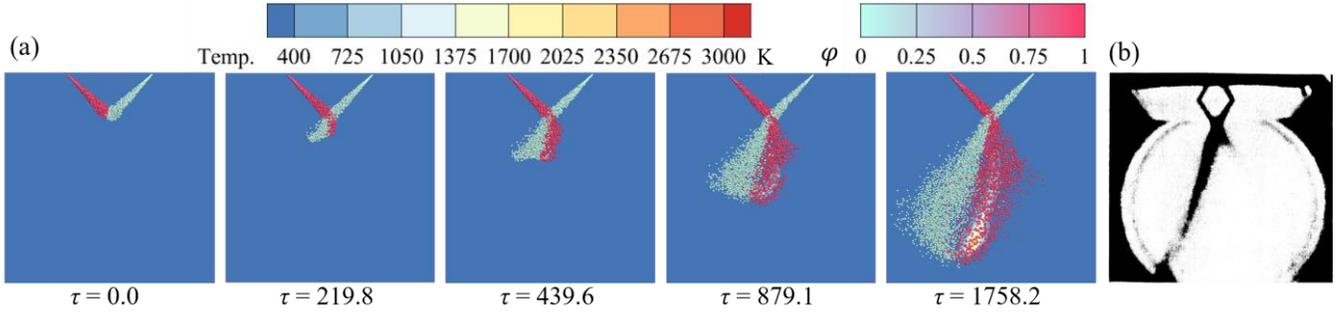

**FIG. 5.** (a) Computational case illustrates the sprays penetrate each other at a smaller $Da_l = 2.84 \times 10^{-3}$ than the benchmark case. (b) Experiment of penetration phenomenon[58].

To demonstrate the existence of the above two modes over wide ranges of Damköhler numbers, we repeated the simulation over a wide range of $Da_g$ and $Da_l$, while $\theta = 90°$ and $L = 4$ mm remain unchanged. As shown in Fig. 6, when $Da_g$ and $Da_l$ are small, the gas-phase and liquid-phase reactions tend to chemically frozen. After the sprays have impinged with each other, evaporated, and combusted, most droplets retain unreactive and pass through each other, causing spray penetration. When both $Da_g$ and $Da_l$ increase simultaneously, the combustion processes of gas-phase and liquid-phase tend toward chemical equilibrium, thereby transforming the combustion mode from penetration to popping. The trend of transition between the two modes is consistent with the experimental results of Lawver et al.[9]. It is interesting to see that the boundary between two mode regimes is close to the unity $Da_g$ and $Da_l$, when the chemical reaction time scales are comparable to the flow time scales. This implies that in actual engine operation, the combustion mode of popping can be avoided by lowering the temperature of the hypergolic propellants to reduce the chemical reaction rate.

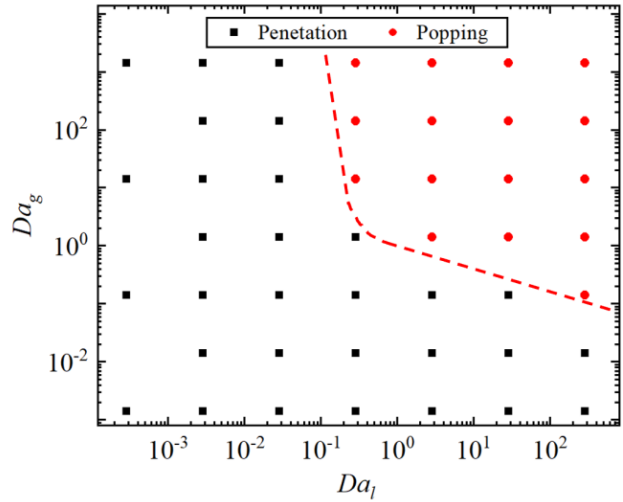

**FIG. 6.** Regime diagram of typical penetration and popping phenomena in the spray impingement combustion process of hypergolic bipropellants.

### 3.2 Physical explanations of popping mode

Fig. 7 displays multiple cyclic processes of the popping phenomenon in benchmark case, with localized magnified images providing further detailed information. At $\tau = 0.0$, the sprays of MMH and NTO impinge upon each other after ejection from the nozzles. Due to the nature of hypergolic propellants, they undergo an intense and exothermic reaction upon contact at the impingement point. At $\tau = 22.0$, the reaction generates a high-temperature and high-pressure zone near the impingement point that causes the sprays to separate. The sprays then disperse at the edges of this high-temperature and high-pressure zone, similar to impacting a high-temperature wall. At $\tau = 44.0$, the high-temperature and high-pressure zone generates a strong thrust, which visibly changes the direction of the droplet's motion toward the surrounding area and accelerates some droplets to exceed their initial velocities at nozzles. After the separation of sprays, the reaction in the impingement point zone subsides due to the depletion of MMH and NTO, leading to a decrease in temperature and pressure. As shown at $\tau = 87.9$, the sprays have a tendency to impinge again. Subsequently, the spray repeats the process of impingement and separation, resulting in the popping phenomenon.

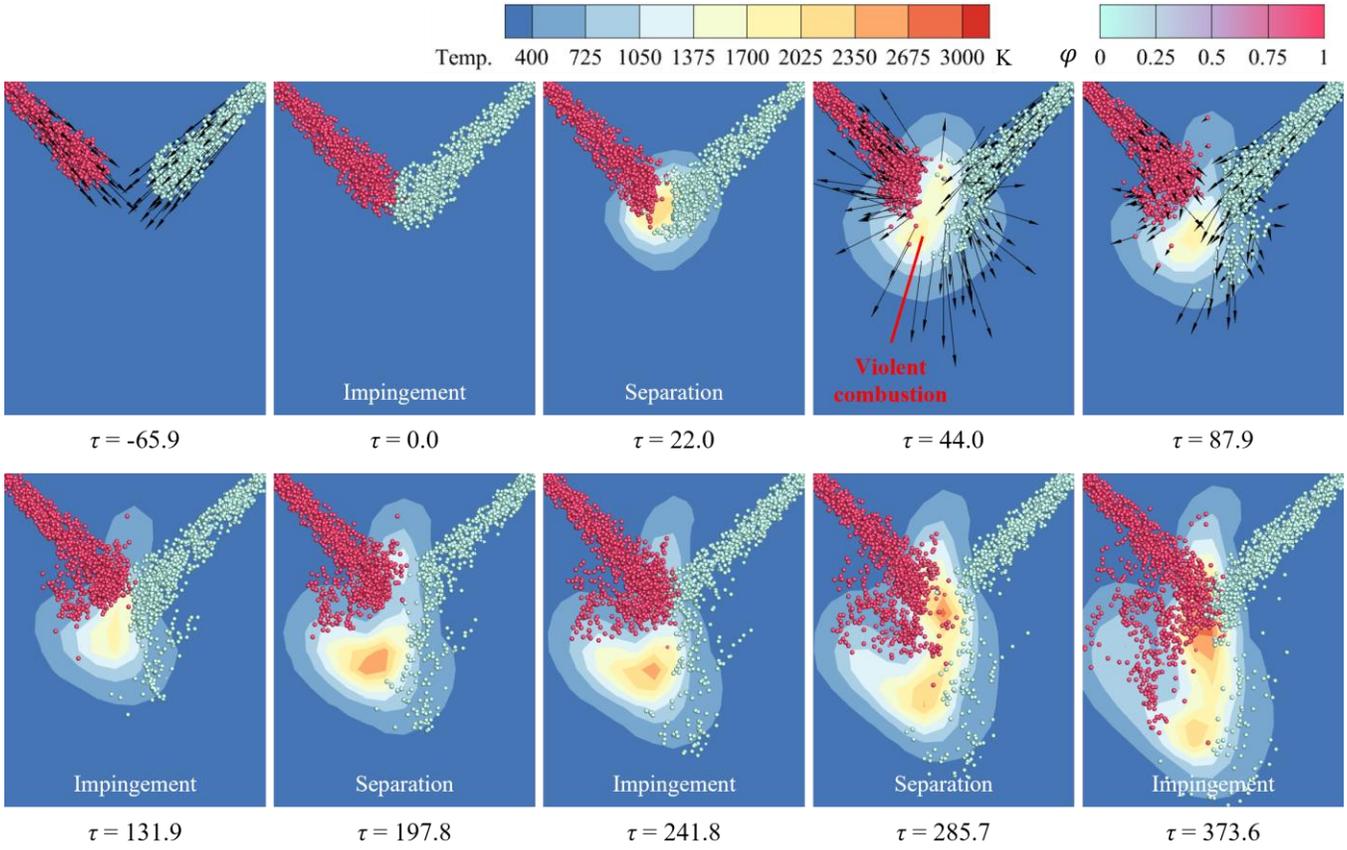

**FIG. 7.** Details of popping phenomenon in the benchmark case ($Da_g$=1.42×10², $Da_l$=2.84×10¹): the sprays of MMH and NTO impinge firstly at $\tau = 0.0$, separate each other at $\tau = 44.0$, and re-impinge at $\tau = 131.9$, continuously cycling through this process. The velocity vectors of some droplets depict their motion direction and magnitude. The temperature field color scale shows gas-phase temperature, while the propellant mass fraction $\varphi$ color scale distinguishes MMH, NTO, and mixed droplets. Mixed droplets are not clearly visible due to their rapid liquid-phase reaction.

Fig. 8 further illustrates the penetration case where the smaller liquid Damköhler number is due to the slower liquid-phase reaction rate. Consequently, there is an insufficiently violent reaction zone around the impingement point to create the high temperature and pressure to attempt to separate the sprays. The temperature rise near the impingement point is small due to the small $Da_l$. Combustion becomes significantly apparent only in the downstream because the gas-phase reactions require some time to accumulate sufficient MMH and NTO vapors. Consequently, heat release from the gas-phase reaction is concentrated along the boundary line separating the two sprays.

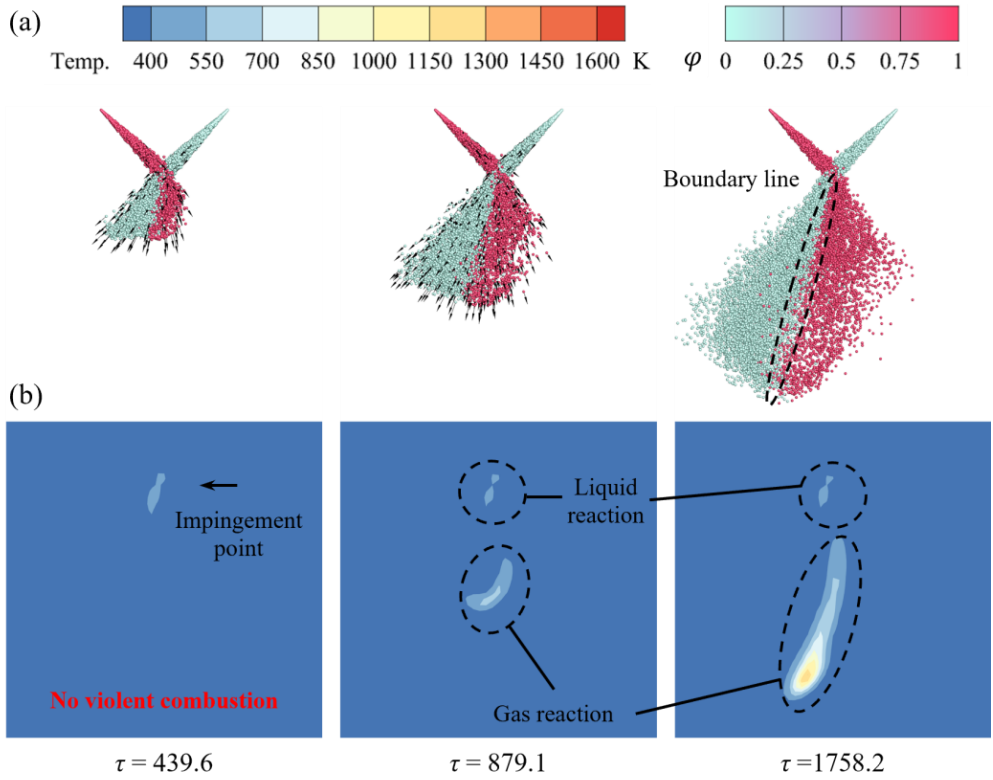

**FIG. 8.** Penetration phenomenon upons spray impingement ($Da_g$=1.42×10$^2$, $Da_l$=2.84×10$^{-3}$): (a) Spray characteristics denoted by moving droplets (b) Gas-phase temperature field.

Some ideal limiting conditions, such as the non-reacting case ($Da_g$=0, $Da_l$=0), the pure gas-phase reaction case ($Da_g$=1.42×10$^7$, $Da_l$=0), and the pure liquid-phase reaction case ($Da_g$=0, $Da_l$=2.84×10$^2$), were compared to reveal the significance of gas-phase and liquid-phase chemical reactions in the popping phenomenon, as shown in Fig. 9. In the non-reacting case, the sprays penetrate each other, with slight vaporization and temperature reduction. The pure gas-phase reaction case is very similar to the penetration case with $Da_l$=2.84×10$^{-3}$. For the pure liquid-phase reaction case, there is a notable difference in spray behavior and the distribution of droplets compared to the non-reacting case and the pure gas-phase reaction case. This difference is attributed to the exothermic liquid-phase reaction in the mixed droplets at the impingement point, which significantly alters droplet characteristics (such as quantity, size, and distribution), thereby affecting the collision and breakup processes of droplets and overall spray behavior. From the spray behavior shown in Fig. 9(a), compared to the non-reacting case and the pure gas-phase reaction case, the pure liquid-phase reaction case shows a sparser spray downstream of the impingement point, due to the intensified evaporation caused by the liquid-phase reaction. As illustrated by the gas-phase temperature field in Fig. 9(b), the non-reaction case absorbs heat from the environment for evaporation without reaction or heat release. The pure gas-phase reaction case generates a high-temperature zone downstream of the impingement point, as MMH and NTO require time to evaporate into the gas. The accumulation of MMH and NTO in the gas phase leads to the formation of a violent combustion zone. In contrast, the pure liquid-phase reaction case only undergoes liquid-phase reactions at the impingement point and along the spray boundary line, failing to replicate the high-temperature zone of the pure gas-phase reaction case. However, it significantly enhances evaporation compared to both the non-reaction case and the pure gas-phase reaction, as shown in Fig. 9(c) and Fig. 9(d), leading to a significant increase in the mass fractions of MMH and NTO in the gas phase. Additionally, the liquid-phase reaction rapidly consumes the mixing droplets, leading to a reduction in the overall droplet count through enhanced evaporation, as shown in Fig. 9(e).

The roles of gas-phase and liquid-phase reactions in producing the popping phenomenon can be summarized as follows. The mixed droplets formed by spray impingement initially undergo the liquid phase reaction. The heat released from the liquid-phase reaction accelerates the evaporation of the spray, converting more MMH and NTO into the gas phase. This conversion facilitates their contact and subsequent gas-phase reactions, generating a large amount of heat. Consequently, the zone of intense reaction, dominated by exothermic gas-phase reactions, is located closer to the spray impingement point. The resulting high temperature and pressure produce thrust to separate the sprays. Upon the depletion of MMH and NTO, the thrust diminishes, and the sprays impinge again, leading to the cyclical popping phenomenon. Both liquid-phase and gas-phase reactions are indispensable to the popping phenomenon.

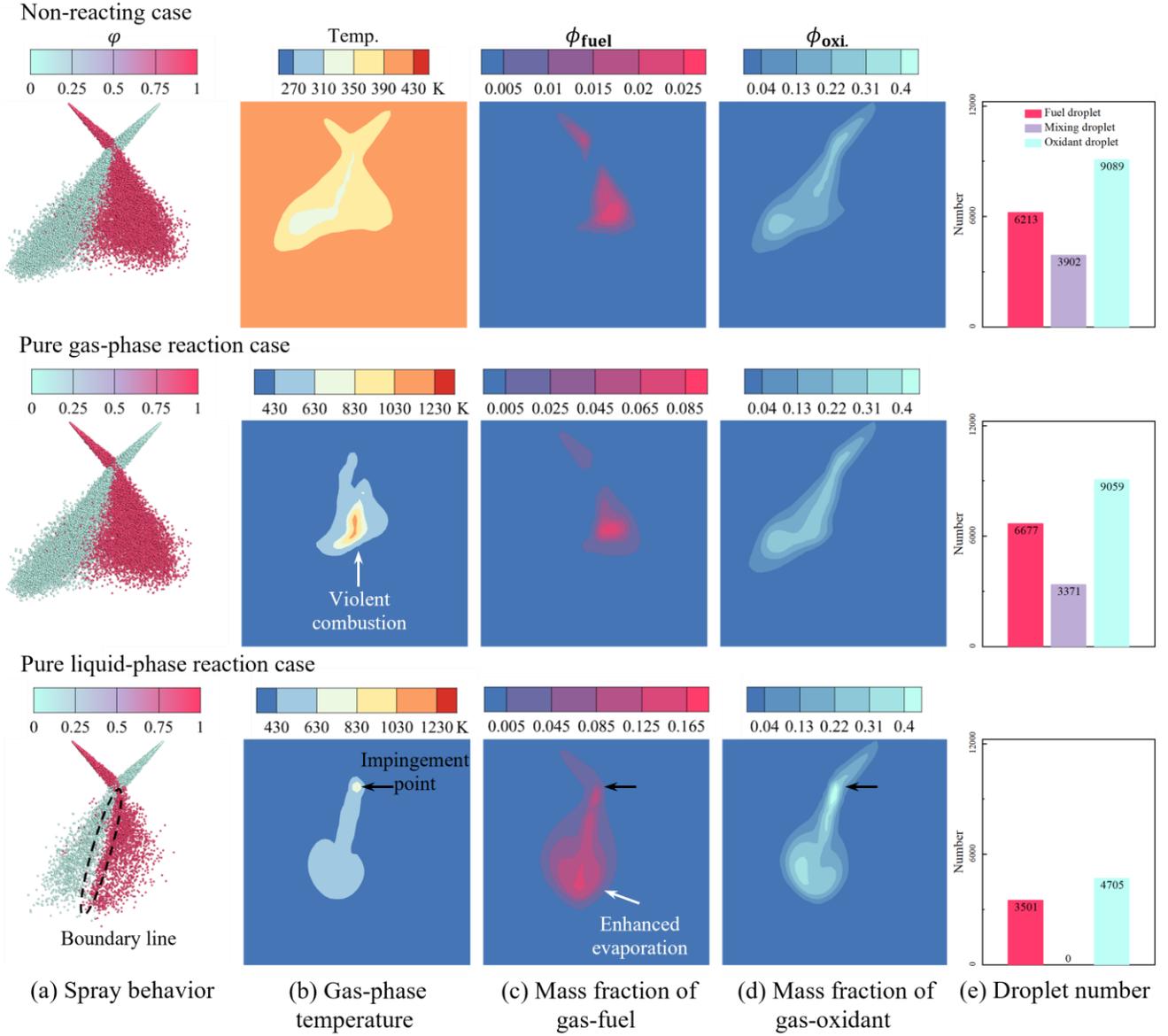

**FIG. 9.** States of the three special cases of at $\tau = 1318.7$: non-reacting case ($Da_g=0$, $Da_l=0$), pure gas-phase reaction case ($Da_g=1.42\times10^7$, $Da_l=0$), pure liquid-phase reaction case ($Da_g=0$, $Da_l=2.84\times10^2$) respectively. (a) Spraying behavior of droplets. (b) Temperature contours of gas-phase. (c) MMH mass fraction contours of in gas-phase. (d) NTO mass fraction contours of gas-phase. (e) Numbers of MMH droplets, mixing droplets, and NTO droplets.

### 3.3 Periodic dynamics of popping mode

To further understand the periodic dynamics of the popping phenomenon, we conducted the Fast Fourier transform (FFT)[59] to obtain the frequency information. Pressure and temperature data are recorded at the spray impingement point, and the number of droplets is counted within an 8mm × 8mm × 10mm cubic box (as shown in Fig. 3). Fig. 10(a), (b), and (c) show the temporal variations of pressure, temperature, and number of droplets for a typical case, respectively. Each exhibits a clear periodicity in the time domain, underscoring the cyclical nature of the popping phenomenon. As shown in Fig. 10(a), the peak pressure of the popping phenomenon in the simulation is consistent with the range reported by Houseman et al.[14], reaching to 50 psi (approximately 3.4 atm) in the experiment. The fluctuations between multiple complete and stable peaks were selected for the FFT with a sampling frequency of 2× $10^{-6}$ Hz to obtain the amplitude-frequency plot, as shown in Fig. 10(d), (e), and (f). The clear dominant frequencies further reveal the periodic nature of the popping phenomenon.

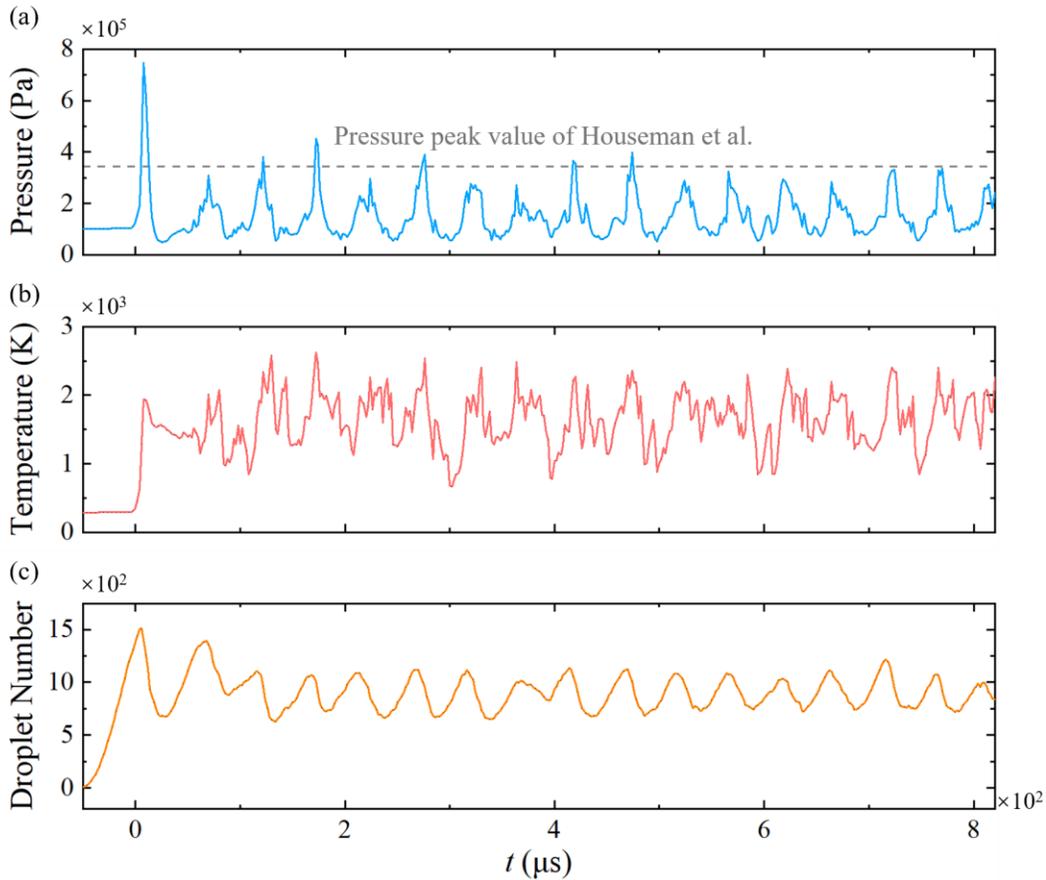

**FIG. 10.** Time domain feature of pressure, temperature, and droplet number in the benchmark case. (a) Temporal variations in pressure at the impingement point. The dashed line represents the peak pressure amplitude for the popping phenomena in the experiment reported by Houseman et al.[14]. (b) Temporal variations in temperature at the impingement point. (c) Temporal variations in droplet number in the 8mm × 8mm × 10mm cubic box near the impingement point.

The Strouhal numbers, $St_P = f_P D/V$, $St_T = f_T D/V$, and $St_N = f_N D/V$, are defined to quantify the periodic variations of pressure, temperature, and number of droplets, respectively, where $f_P$, $f_T$, and $f_N$ are the dominant frequencies of pressure, temperature and droplet number, respectively. As illustrated in Fig. 11, $St_P$, $St_T$ and $St_N$ are remarkably similar across all popping combustion mode cases, with values around $St_P \sim St_T \sim St_N \sim O(1 \times 10^{-2})$, which is consistent with the range reported by Campbell[16]. It is interesting to see that all the Strouhal numbers show a small variation with $Da_g$ and $Da_l$, and they may be mainly influenced by the injection angle, the spray mass flow rates, and the ambient conditions. These factors will be the focus of our future research.

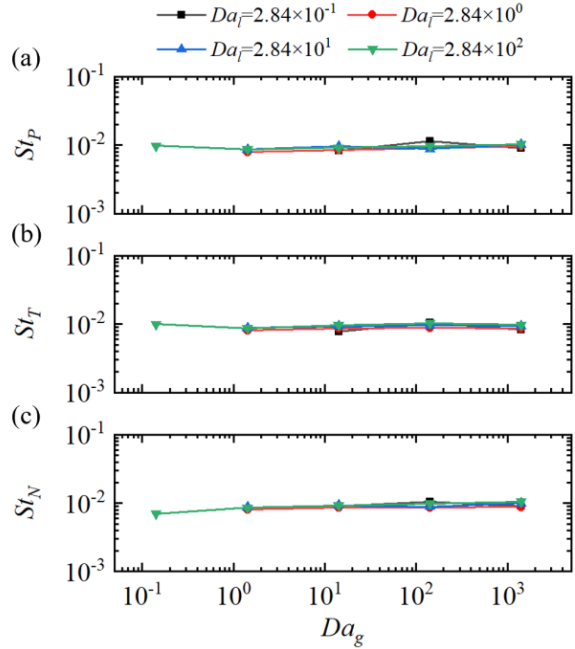

**FIG. 11.** Diagram of $St - Da_g$ at the range of $Da_l$ from $2.84 \times 10^{-1}$ to $2.84 \times 10^2$, for popping mode. (a-c) The frequency is obtained from the pressure at the impingement point, the temperature at the impingement point, and the droplet number in an 8mm × 8mm × 10mm cubic box near the impingement point.

## 4. Concluding Remarks

This paper employed a Eulerian-Lagrangian approach to simulate the ignition process of hypergolic bipropellants of MMH/NTO. Of particular consideration is the liquid droplet inter-mixing process followed by liquid-phase reactions based on an extended spray equation. Different combustion modes were computationally realized, their dynamics are analyzed by the heat and flow field information, and the roles of gas-phase and liquid-phase reactions in determining these combustion modes were clarified. The important conclusions are summarized as follows:

The popping phenomenon is computationally reproduced over wide ranges of Damköhler numbers of liquid- and gas-phase reactions for the first time, and the results were both qualitatively and quantitatively validated against previous experimental studies. The two sprays impinge firstly, separate by thrust from the high temperature and pressure combustion zone at the impingement point, and re-impinge as the depletion of MMH and NTO, forming a cyclical popping phenomenon. The popping mode is significantly influenced by the Damköhler numbers. As the Damköhler numbers decrease, meaning the characteristic flow time becomes shorter than the gas-phase and liquid-phase chemical reaction times, the popping mode shifts to the penetration mode, where the sprays penetrate each other upon impinge without separation. In the popping mode, the exothermic liquid-phase reaction enhances the propellant evaporation, which in turn promotes the gas-phase reaction around the spray impingement point. The periodic frequency of the popping phenomenon, characterized by fluctuations in pressure, temperature, and droplet number, is minimally influenced by the Damköhler numbers and may be primarily controlled by other factors, meriting further investigation in future studies.

This paper contributes to a new understanding of the hypergolic ignition of MMH/NTO in rocket engines. By controlling propellant temperature and flow rate to adjust the Damköhler numbers, the phenomenon of off-design popping can be avoided, aiding in the design of liquid rocket motors to maintain combustion efficiency and stability. Future computational work will adopt more sophisticated combustion and turbulence models and utilize more detailed mechanisms of both liquid- and gas-phase reactions of MMH/NOT. In addition, the effects of droplet breakup, droplet collision, droplet inter-mixing, and vaporization of multicomponent droplets on the liquid-phase reactions of droplets need to be further investigated.

## CRediT authorship contribution statement

**Jinyang Wang:** Writing – original draft, Validation, Methodology, Visualization, Software, Formal analysis. **Kai Sun:** Writing – review & editing, Formal analysis, Supervision, **Tianyou Wang:** Resources, Funding acquisition, Supervision. **Peng Zhang:** Writing – review & editing, Supervision, Resources, Funding acquisition, Formal analysis, Conceptualization.

## Acknowledgment

This work was financially supported by the National Natural Science Foundation of China (No. 52176134 and 51921004). The work at the City University of Hong Kong was additionally supported by grants from the Research Grants Council of the Hong Kong Special Administrative Region, China (Project No. CityU 15222421 and CityU 15218820). The authors gratefully acknowledge Dr. Tao Yang for their insightful advice for the data processing and to Dr. Qiankun Zhang for his suggestion for numerical simulation.

## Declaration of competing interest

The authors have no conflicts to disclose.

## Data availability

The data that support the findings of this study are available from the corresponding author upon reasonable request.

# References

[1] Yuan T, Chen C, Huang B, Tang M, Chen Y-T. The impinging-type injector design of MMH/NTO liquid rocket engine. 48th AIAA/ASME/SAE/ASEE Joint Propulsion Conference & Exhibit, Atlanta, Georgia: American Institute of Aeronautics and Astronautics; 2012. https://doi.org/10.2514/6.2012-3745.

[2] Sutton GP, Biblarz O. Rocket propulsion elements. John Wiley & Sons; 2016.

[3] Sutton GP. History of liquid propellant rocket engines in the United States. Journal of Propulsion and Power 2003;19:978–1007.

[4] Davis SM, Yilmaz N. Advances in hypergolic propellants: Ignition, hydrazine, and hydrogen peroxide research. Advances in Aerospace Engineering 2014;2014.

[5] Zung LB, Tkachencko EA, Breen BP. A basic study on the ignition of hypergolic liquid propellants. NASA Technical Report No CR--100387 1968.

[6] Trinh HP, Burnside C, Williams H. Assessment of MON-25/MMH Propellant System for Deep-Space Engines. International Astronautical Congress (IAC), 2019.

[7] Houseman J, Kushida R. Criteria for separation of impinging streams of hypergolic propellants. FALL MEETING, COMBUSTION INST.; 1967.

[8] Lawver BR. High performance N2O4/amine elements: Blowapart. 1977.

[9] Lawver BR, Breen BP. Hypergolic Stream Impingement Phenomena-Nitrogen Tetroxide/Hydrazine. NASA Contractor Report 1968;72444:25.

[10] Somogyi D, Feiler CE. Liquid-Phase Heat-Release Rates of the Systems Hydrazine-Nitric Acid and Unsymmetrical Dimethylhydrazine-Nitric Acid. National Aeronautics and Space Administration; 1960.

[11] Burrows MC. Mixing and reaction studies of hydrazine and nitrogen tetroxide using photographic and spectral techniques. National Aeronautics and Space Administration; 1968.

[12] Dennis JD, Son SF, Pourpoint TL. Critical Ignition Criteria for Monomethylhydrazine and Red Fuming Nitric Acid. Journal of Propulsion and Power 2015;31:1184–92. https://doi.org/10.2514/1.B35541.

[13] Tani H, Daimon Y, Sasaki M, Matsuura Y. Atomization and hypergolic reactions of impinging streams of monomethylhydrazine and dinitrogen tetroxide. Combustion and Flame 2017;185:142–51. https://doi.org/10.1016/j.combustflame.2017.07.005.

[14] Houseman J, Lee A. Popping Phenomena with the Hydrazine Nitrogen-Tetroxide Propellant System. Journal of Spacecraft and Rockets 1972;9:678–82. https://doi.org/10.2514/3.61775.

[15] Nurick WH, Cordill JD. FINAL REPORT REACTIVE STREAM SEPARATION PHOTOGRAPHY n.d.

[16] Campbell DT, Clapp SD, Proffit RL, Cline GL. Reactive Stream Separation Photography. AIAA Journal 1971;9:1832–6. https://doi.org/10.2514/3.49985.

[17] Houseman J. Optimum mixing of hypergolic propellants in an unlike doublet injector element. AIAA Journal 1970;8:597–9. https://doi.org/10.2514/3.5724.

[18] Zhang D, Zhang P, Yuan Y, Zhang T. Hypergolic ignition by head-on collision of N, N, N′, N′−tetramethylethylenediamine and white fuming nitric acid droplets. Combustion and Flame 2016;173:276–87.

[19] Zhang D, Yu D, Zhang P, Yuan Y, Yue L, Zhang T, et al. Hypergolic ignition modulated by head-on collision, intermixing and convective cooling of binary droplets with varying sizes. International Journal of Heat and Mass Transfer 2019;139:475–81.

[20] Zhang D, He C, Zhang P, Tang C. Mass interminglement and hypergolic ignition of TMEDA and WFNA droplets by off-center collision. Combustion and Flame 2018;197:276–89.

[21] FLETCHER EA, MORRELL G. Ignition in liquid propellant rocket engines. Progress in Combustion Science and

Technology, Elsevier; 1960, p. 183–215.

[22] He C, He Z, Zhang P. Droplet collision of hypergolic propellants. Droplet 2024;3:e116. https://doi.org/10.1002/dro2.116.

[23] Knab O, Preclik D, Estublier D. Flow field prediction within liquid film cooled combustion chambers of storable bi-propellant rocket engines. 34th AIAA/ASME/SAE/ASEE Joint Propulsion Conference and Exhibit, Cleveland,OH,U.S.A.: American Institute of Aeronautics and Astronautics; 1998. https://doi.org/10.2514/6.1998-3370.

[24] Wei Q, Liang G. Coupled Lagrangian impingement spray model for doublet impinging injectors under liquid rocket engine operating conditions. Chinese Journal of Aeronautics 2017;30:1391–406. https://doi.org/10.1016/j.cja.2017.06.011.

[25] Lee Y-W, Jiang T-L. Effects of Fuel Impingement-Cooling on the Combustion Flow in a Small Bipropellant Liquid Rocket Thruster. J Mech 2015;31:161–70. https://doi.org/10.1017/jmech.2014.81.

[26] Ohminami K, Ogawa H, Uesugi K. Numerical Bipropellant Thruster Simulation with Hydrazine and NTO Reduced Kinetic Reaction Model. 47th AIAA Aerospace Sciences Meeting including The New Horizons Forum and Aerospace Exposition, Orlando, Florida: American Institute of Aeronautics and Astronautics; 2009. https://doi.org/10.2514/6.2009-452.

[27] Lee KH. Numerical simulation on thermal and mass diffusion of MMH–NTO bipropellant thruster plume flow using global kinetic reaction model. Aerospace Science and Technology 2019;93:104882. https://doi.org/10.1016/j.ast.2018.11.056.

[28] Chu W, Guo K, Tong Y, Li X, Nie W. Numerical analysis of self-excited tangential combustion instability for an MMH/NTO rocket combustor. Proceedings of the Combustion Institute 2023;39:5053–61. https://doi.org/10.1016/j.proci.2022.07.249.

[29] Wei Q, Liang G. Numerical Simulation of Ignition Process for the Monomethyl Hydrazine–Nitrogen Tetroxide Thrusters. Journal of Propulsion and Power 2019;35:704–19. https://doi.org/10.2514/1.B37136.

[30] Liu W-G, Wang S, Dasgupta S, Thynell ST, Goddard WA, Zybin S, et al. Experimental and quantum mechanics investigations of early reactions of monomethylhydrazine with mixtures of NO2 and N2O4. Combustion and Flame 2013;160:970–81. https://doi.org/10.1016/j.combustflame.2013.01.012.

[31] Zhang Q, Zhang P, Chi Y, Yang T, Zhu J, Lu X. Eulerian-Lagrangian simulation and validating experiment of n-butanol/biodiesel dual-fuel impinging sprays. Fuel 2023;350:128761. https://doi.org/10.1016/j.fuel.2023.128761.

[32] Zhang Q, Hu Z, Zhu J, Zhang P, Lu X. Experimental and numerical study on mixture concentration distribution and ignition characteristics of biodiesel/n-butanol dual-fuel spray impingement in inert and combustion environments. Fuel 2024;368:131592. https://doi.org/10.1016/j.fuel.2024.131592.

[33] Amsden AA, O'Rourke PJ, Butler TD. KIVA-II: A computer program for chemically reactive flows with sprays. Los Alamos National Lab.(LANL), Los Alamos, NM (United States); 1989.

[34] Amsden AA. KIVA-3: A KIVA program with block-structured mesh for complex geometries. Los Alamos National Lab., NM (United States); 1993.

[35] Amsden A. KIVA3V. A Block-Structured KIVA Program for Engines with Vertical or Canted Valves. Los Alamos National Lab.(LANL), Los Alamos, NM (United States); 1997.

[36] Wang Y, Lee WG, Reitz RD, Diwakar R. Numerical simulation of diesel sprays using an eulerian-lagrangian spray and atomization (ELSA) model coupled with nozzle flow. SAE Technical Paper; 2011.

[37] Li Y, Cai Y, Jia M, Wang Y, Su X, Li L. A full-parameter computational optimization of both injection parameters and injector layouts for a methanol/diesel dual-fuel direct injection compression ignition engine. Fuel 2024;369:131733.

[38] Rachner M, Becker J, Hassa C, Doerr T. Modelling of the atomization of a plain liquid fuel jet in crossflow at gas turbine conditions. Aerospace Science and Technology 2002;6:495–506.

[39] Pai GM, Subramaniam S. ACCURATE NUMERICAL SOLUTION OF THE SPRAY EQUATION USING PARTICLE METHODS. Atomiz Spr 2006;16:159–94. https://doi.org/10.1615/AtomizSpr.v16.i2.30.

[40] Williams FA. Spray combustion and atomization. The Physics of Fluids 1958;1:541–5.

[41] Subramaniam S. Statistical modeling of sprays using the droplet distribution function. Physics of Fluids 2001;13:624–42. https://doi.org/10.1063/1.1344893.

[42] Schröder A, Schanz D. 3D Lagrangian Particle Tracking in Fluid Mechanics. Annual Review of Fluid Mechanics 2023;55:511–40. https://doi.org/10.1146/annurev-fluid-031822-041721.

[43] Pope SB. PDF methods for turbulent reactive flows. Progress in Energy and Combustion Science 1985;11:119–92. https://doi.org/10.1016/0360-1285(85)90002-4.

[44] Subramaniam S. Lagrangian–Eulerian methods for multiphase flows. Progress in Energy and Combustion Science 2013;39:215–45. https://doi.org/10.1016/j.pecs.2012.10.003.

[45] O'Rourke PJ, Amsden AA. The TAB method for numerical calculation of spray droplet breakup. SAE Technical Paper; 1987.

[46] O'ROURKE PJ. Collective drop effects on vaporizing liquid sprays. Princeton University, 1981.

[47] Sun K, Jia F, Zhang P, Shu L, Wang T. Marangoni Effect in Bipropellant Droplet Mixing during Hypergolic Ignition. Phys Rev Applied 2021;15:034076. https://doi.org/10.1103/PhysRevApplied.15.034076.

[48] Jia F, Peng X, Wang J, Wang T, Sun K. Marangoni-driven spreading of a droplet on a miscible thin liquid layer. Journal of Colloid and Interface Science 2024;658:617–26. https://doi.org/10.1016/j.jcis.2023.12.092.

[49] Dreeben TD, Pope SB. Nonparametric estimation of mean fields with application to particle methods for turbulent flows. Sibley School of Mechanical and Aerospace Engineering, Cornell University, Technical Report No FDA 1992:92–13.

[50] Sirignano WA. Fuel droplet vaporization and spray combustion theory. Progress in Energy and Combustion Science 1983;9:291–322. https://doi.org/10.1016/0360-1285(83)90011-4.

[51] Law CK. Recent advances in droplet vaporization and combustion. Progress in Energy and Combustion Science 1982;8:171–201. https://doi.org/10.1016/0360-1285(82)90011-9.

[52] Law CK. Combustion physics. Cambridge university press; 2010.

[53] Turns SR. Introduction to combustion. vol. 287. McGraw-Hill Companies New York, NY, USA; 1996.

[54] White JR, Zung LB. Combustion process of impinging hypergolic propellants. 1971.

[55] Zhang Z, Chi Y, Shang L, Zhang P, Zhao Z. On the role of droplet bouncing in modeling impinging sprays under elevated pressures. International Journal of Heat and Mass Transfer 2016;102:657–68. https://doi.org/10.1016/j.ijheatmasstransfer.2016.06.052.

[56] Kim S, Lee DJ, Lee CS. Modeling of binary droplet collisions for application to inter-impingement sprays. International Journal of Multiphase Flow 2009;35:533–49. https://doi.org/10.1016/j.ijmultiphaseflow.2009.02.010.

[57] Suo S, Jia M, Liu H, Wang T. Development of a New Hybrid Stochastic/Trajectory Droplet Collision Model for Spray Simulations in Internal Combustion Engines. International Journal of Multiphase Flow 2021;137:103581. https://doi.org/10.1016/j.ijmultiphaseflow.2021.103581.

[58] Falk AY. High performance N2O4/amine elements. 1976.

[59] Stein EM, Shakarchi R. Fourier analysis: an introduction. vol. 1. Princeton University Press; 2011.